\documentclass{IEEEtran}

\pdfoutput=1

\usepackage[T1]{fontenc}
\usepackage[utf8]{inputenc}
\usepackage{stix}
\usepackage[ruled,vlined]{algorithm2e}
\usepackage{amsthm}
\usepackage{blkarray}
\usepackage{bm}
\usepackage{booktabs}
\usepackage{cite}
\usepackage{mathtools}
\usepackage{microtype}
\usepackage{multirow}
\usepackage{relsize}
\usepackage{xcolor}
\usepackage{hyperref}

\hypersetup{
    pdftitle={Histogram Specification by Assignment of Optimal Unique Values},
    pdfauthor={Vítor Saraiva Ramos; Luiz Felipe de Queiroz Silveira; Luiz Gonzaga de Queiroz Silveira Júnior},
    pdfkeywords={Algorithms; Data preprocessing; Data processing; Histograms; Optimization; Sorting},
    colorlinks
}

\theoremstyle{definition}
\newtheorem{definition}{Definition}

\newcommand*{\abs}[1]{\lvert#1\rvert}
\newcommand*{\norm}[1]{\lVert#1\rVert}
\newcommand*{\set}[1]{\lbrace#1\rbrace}
\newcommand*{\jj}{{j+1}-1}
\newcommand*{\jjj}[1]{#1_{j}\dots#1_\jj}
\newcommand*{\mailto}[2]{\href{mailto:#1@#2}{#1@#2}}

\SetCommentSty{AlCommentSty}

\DeclareMathOperator{\argmin}{argmin}
\DeclareMathOperator{\argsort}{argsort}
\DeclareMathOperator{\card}{card}
\DeclareMathOperator{\counts}{counts}
\DeclareMathOperator{\cumsum}{cumsum}
\DeclareMathOperator{\sort}{sort}
\DeclareMathOperator{\unique}{unique}

\begin{document}

\title{Histogram Specification\\%
    by Assignment of Optimal Unique Values%
}

\author{Vítor~S.~Ramos,~\IEEEmembership{Student Member,~IEEE,} Luiz~Felipe~de~Q.~Silveira,~\IEEEmembership{Member,~IEEE,} and~Luiz~Gonzaga~de~Q.~Silveira~Júnior,~\IEEEmembership{Member,~IEEE}%
    \thanks{V. S. Ramos is with the Electrical and Computer Engineering Graduate Program, Federal University of Rio Grande do Norte, Natal, Brazil (e-mail: \mailto{vitorsr}{ufrn.edu.br}). L. F. de Q. Silveira is with the Computer Engineering and Automation Department, Federal University of Rio Grande do Norte, Natal, Brazil (e-mail: \mailto{lfelipe}{dca.ufrn.br}). L. G. de Q. Silveira Júnior is with the Communications Engineering Department, Federal University of Rio Grande do Norte, Natal, Brazil (e-mail: \mailto{junior}{ct.ufrn.br}).}%
    \thanks{This study was financed in part by the \textit{Coordenação de Aperfeiçoamento de Pessoal de Nível Superior}---\textit{Brasil} (CAPES)---Finance Code 001.}%
}

\maketitle

\begin{abstract}
    In this paper, we propose two novel algorithms for histogram specification and quantile transformation of data without local information.
    These are core techniques that can serve as building blocks for applications that require specifying the sample distribution of a given set of data.
    Histogram specification is best known for its image enhancement applications, whereas quantile transformation is typically employed in data preprocessing for data normalization.
    In signal processing, methods often require temporal or spatial information; in data preprocessing, methods work by interpolation or by approximation, drawing from results in computational statistics, and have a trade-off between speed and quality.
    It is nontrivial to accommodate for cases that do not have local information (e.g., tabular data) while also providing a fast, exact solution.
    For that, we take up a concept in image processing called group mapping law and propose an extension.
    The proposed extension allows us to formulate a convex functional where we look for the best approximation between the output unique values and the reference histogram.
    Then, we apply the ordered assignment solution, a result in optimal transport, to reconstruct the output from the optimal unique values.
    Two sets of results show the effectiveness of the proposed algorithms when compared to traditional and state-of-the-art methods.
    The proposed algorithms are fast, exact, and least $p$-norm optimal.
    Further, we define the algorithms as generic data processing methods.
    Thus, contributions from this paper can be easily incorporated in applications spanning many disciplines, especially in applied data science.
\end{abstract}


\section{Introduction}

\IEEEPARstart{T}{he} technique of histogram equalization first appeared in the context of real-time image enhancement for cockpit display systems \cite{Ketcham1974}.
Histogram equalization refers to the task of adjusting the histogram of input data such that it follows a uniform distribution.
It is a subset of histogram specification, where the task is to adjust an input histogram such that it best approximates the histogram of a given reference.
The latter technique is also known as histogram matching, modeling, or transfer \cite{Zhang1992, Balado2018}.

These techniques are notoriously featured in image enhancement works \cite{Zhang1992, Coltuc2006, Thomas2011, Nikolova2014, Balado2018}.
That is because numeric values in images refer to intensity levels, and intensity transformations are global mapping operations that are visually interpretable as contrast modifications \cite{Bauer2016}.
These techniques play a key role in intensity matching.
For instance, we can find applications in microscopy, to compensate light attenuation \cite{Stanciu2009}; in stereoscopic cinema, to match colors in twin cameras \cite{Bertalmio2014}; in ophthalmic imaging, to extend signal strength in tomography images \cite{Chen2015}; and in image enhancement, to remove highlights from a single image \cite{Ramos2020}.

Moreover, although histogram equalization is most closely related to image processing, it also appears in data engineering, known in practice under the name of quantile normalization.
In this context, we highlight the very widespread technique of quantile normalization of high-density oligonucleotide array data, as a means of data preprocessing \cite{Bolstad2003}.

Our target use case is applied data science, where data is most commonly structured in tabular form.
In specific, we regard data preprocessing, which includes data normalization \cite{Garcia2016, Kuhn2019}.
In \cite{Xin2018}, Xin \textit{et al.} note that most fields employ normalization algorithms in the data preprocessing step, and in \cite{Munson2012}, Munson highlights that researchers spend approximately a third of their time with data preprocessing.
Likewise, automated machine learning methods and systems typically evaluate several normalization techniques in their first steps \cite{Dirac2018, Hutter2019}, thus reaffirming the importance of this practice.

That said, much like histogram equalization, one of the subset applications of histogram specification is to perform data normalization by quantile transformation.
However, current implementations are based on sample quantile estimation \cite{Hyndman1996}, followed by interpolating the cumulative distribution function (CDF) estimate given by the quantiles; or by approximate quantile computation \cite{Chen2020}, followed by evaluating the approximate CDF.
Current implementations may be slow or inexact depending on the number of quantiles estimated or on the quantile approximation algorithm employed.

On a higher level, there are several works related to the problem at hand.
First, the assignment---or matching---problem is long known to the optimal transport literature.
Fast solutions are known for $p=1$ but do not regard conflict-free (bijective) assignments \cite{Karp1975, Werman1986, Peyre2019}.
The shortcomings of assignments that disregard conflicts are known to image processing literature (see, e.g., \cite[Fig. 1]{Nikolova2014}).
Second, many fast solutions have been proposed for the exact histogram specification problem in image processing \cite{Zhang1992, Coltuc2006, Thomas2011, Nikolova2014, Balado2018}.
However, they only treat integers and they also require local information (i.e., they consider temporal or spatial structure in the data).
None of the proposed methods, to the best of our knowledge, are apt to transform tabular data without major, nontrivial modifications.

In summary, fast algorithms yielding optimal data transformations that do not require local information are desirable.
Thus, we present in this paper contributions towards fast least $p$-norm algorithms for histogram specification of data without local information.

\subsection{Contributions}

The contributions from this paper are as follows.
Firstly, we construct a special matrix $\bm{A}$ to preserve the mapping bijectivity, extending the group mapping law \cite{Zhang1992}.
This matrix allows us to express an ordered array in function of its unique values.
Secondly, we express the problem as a convex formulation of the best approximation between the output unique values and the reference.
We reconstruct the output by assignment of optimal unique values.
Thirdly, we present a generic histogram specification algorithm fit for tabular data.
Lastly, we incorporate results from computational statistics to present a fast, exact quantile transformer algorithm.

\section{Definitions}

Initially, we must establish definitions for some of the mathematical operators used throughout the manuscript.
We will reuse the symbols used in the definitions without loss of precision.
Refer also to \autoref{tab:definitions} for a list of symbols used.

In regard to notation, we will follow ISO 80000-2 rules to typeset mathematics \cite{ISO80000}.
That is by choice, to be able to unify notation across multidisciplinary works we reference.
Summarily, italic bold symbols typeset with uppercase letters represent multirow or multicolumn matrices (e.g., $\bm{A}$); italic bold lowercase letters represent vectors (e.g., $\bm{x}$); and italic lowercase letters represent scalars (e.g., $n$).
In this work, we additionally employ italic bold lowercase Greek letters (e.g., $\bm{\phi}$) to refer to vectors of indexing variables (i.e., unsigned integers).
We denote an indexing operation via subscript: $x_{\phi_{0}}$ denotes the element $\phi_{0}$ of $\bm{x}$, where similarly $\phi_{0}$ denotes the element 0 (first) of $\bm{\phi}$.

We will consider an input array $\bm{x}=[x_{0}, x_{1}, \dots, x_{n-1}]^{\top}$ with $x_{i}\in\mathcal{X}$ and $n$ a positive integer.
The set $\mathcal{X}$ can be any numeric (e.g., reals) or ordinal categorical set that is possible to totally order \cite{Knuth1998}.
We will not use nor operate directly on proper values $x_{i}$, therefore we place no restrictions on $x_{i}$ other than being able to sort it and obtain its unique values.

In contrast to available technical literature, exact histogram specification methods operate with integer transformations \cite{Zhang1992, Coltuc2006, Thomas2011, Nikolova2014, Balado2018}, and quantile estimators in statistical packages require numerical values to perform interpolation \cite{Hyndman1996}.
Our input is thus broader in this regard.

\begin{definition}
    Let $\argsort(\cdot)$ be a function that returns an array of indices that sort its argument.
    For an input $\bm{x}$, $\argsort(\bm{x})=[\phi_{0}, \phi_{1}, \dots, \phi_{n-1}]^{\top}$ s.t. $x_{\phi_{0}} \le x_{\phi_{1}} \le \dots \le x_{\phi_{n-1}}$ and $\phi_{i} \neq \phi_{i^{\prime}}$ for all $i \neq i^{\prime}$.
    In addition, $x_{\bm{\phi}}:=\sort(\bm{x})$.
\end{definition}

In \cite{Balado2018}, $\argsort(\bm{x})$ is defined via a permutation matrix $\bm{\mathit{\Pi}}_{\bm{x}}$, such that the multiplication of this matrix with an input array $\bm{x}$ yields sorted values.
These definitions are equivalent.
However, since we will make use of the indices, we prefer the above definition.

\begin{table}
    \centering
    \caption{List of Symbols}
    \label{tab:definitions}
    \begin{tabular}{@{}ccc@{}}
        \toprule
        Notation      & Domain                     & Description                                     \\\midrule
        $i$           & $\mathbb{Z}_{\ge 0}$       & Indexing variable for $n$-arrays                \\
        $j$           & $\mathbb{Z}_{\ge 0}$       & Indexing variable for $m$-arrays                \\
        $m$           & $\mathbb{Z}_{\ge 1}$       & Length of $\bm{e}$                              \\
        $n$           & $\mathbb{Z}_{\ge 1}$       & Length of $\bm{x}$                              \\
        $p$           & $\mathbb{R}_{\ge 1}$       & $p$ in $\ell^{p}$ norm (also called $p$-norm)   \\
        $\bm{e}$      & $\mathcal{E}^{m}$          & Sorted unique values of the input $\bm{x}$      \\
        $\bm{u}$      & $\mathbb{R}^{m}$           & Sorted unique values of the output $\bm{y}$     \\
        $\bm{v}$      & $\mathbb{R}^{n}$           & Reference array                                 \\
        $\bm{x}$      & $\mathcal{X}^{n}$          & Input array                                     \\
        $\bm{y}$      & $\mathbb{R}^{n}$           & Output array                                    \\
        $\bm{A}$      & $\set{0, 1}^{n \times m}$  & Group mapping law matrix of $\bm{x}$            \\
        $\alpha$      & $\mathbb{R}$               & Uniform specification parameter (1)             \\
        $\beta$       & $\mathbb{R}$               & Uniform specification parameter (2)             \\
        $\gamma$      & $\mathbb{R}$               & Uniform specification parameter (3)             \\
        $\bm{\phi}$   & $\mathbb{Z}_{\ge 0}^{n}$   & Indices (arguments) that sort $\bm{x}$          \\
        $\bm{\psi}$   & $\mathbb{Z}_{\ge 1}^{m}$   & Counts of each sorted unique value of $\bm{x}$  \\
        $\bm{\omega}$ & $\mathbb{Z}_{\ge 0}^{m+1}$ & Indices of the unique value groups of $\bm{Ae}$ \\
        \bottomrule
    \end{tabular}
\end{table}

\begin{definition}
    Let $\unique(\cdot)$ be a function that returns an array with the sorted unique values of its argument.
    For an input $\bm{x}$, let $\mathcal{E}\subseteq\mathcal{X}$ be the set of its unique values, that is, $\mathcal{E}=\bigcup_{i=0}^{n-1}\set{x_{i}}$.
    $\mathcal{E}$ is not known \textit{a priori}.
    Then, $\unique(\bm{x})=[e_{0}, e_{1}, \dots, e_{m-1}]^{\top}$, where $e_{j}\in\mathcal{E}$ for all $j$, $e_{0} < e_{1} < \dots < e_{m-1}$, and $m=\card\mathcal{E}$.
    For instance, we have $e_{0}=\min\mathcal{E}$ and $e_{m-1}=\max\mathcal{E}$.
\end{definition}

\begin{definition}
    Let $\counts(\cdot)$ be a function that returns an array with the counts for each unique value in an input array.
    For an input $\bm{x}$, $\counts(\bm{x})=[\psi_{0}, \psi_{1}, \dots, \psi_{m-1}]^{\top}$.
    Each element $\psi_{j}$ is defined as $\psi_{j}=\sum_{i=0}^{n-1} \mathbb{1}_{x_{i}=e_{j}}$, for $j=0 \dots m-1$, where $\bm{e}=\unique(\bm{x})$, and $\mathbb{1}_{x_{i}=e_{j}}$ is one if the logical expression $x_{i}=e_{j}$ is true and zero otherwise.
\end{definition}

To improve reproducibility, the naming and the above definitions of $\unique(\cdot)$ and $\counts(\cdot)$ are consistent with the latest stable release of the numpy.unique function \cite{Harris2020}.
In practice, the above-defined mathematical operators are implemented by algorithms with underlying $O(n \log n)$ time complexity.

\section{Mathematical Preliminaries}

In this section, we present the mathematical foundation upon which we will be able to propose the two algorithms.

\subsection{Problem Statement}

In this paper, we are concerned with the problem of applying the forward map $T:\mathcal{X}^{n} \times \mathcal{V}^{n} \rightarrow \mathcal{Y}^{n}; (\bm{x}, \bm{v}) \mapsto \bm{y}$ such that $d(\mathsf{h}_{\bm{y}}, \mathsf{h}_{\bm{v}})$ is minimized over a distance $d$ for histograms $\mathsf{h}_{\bm{y}}$ and $\mathsf{h}_{\bm{v}}$, and that $\counts(\bm{y})=\counts(\bm{x})$.
The latter condition is such to provide a bijective transformation.
We take input $\bm{x}$ and reference $\bm{v}$ to have same dimensions, and we consider the distance to be the $\ell^{p}$ norm between the sorted output and reference vectors, $d=\norm{\sort(\bm{y})-\sort(\bm{v})}_{p}$.
This distance is the $p$-Wasserstein distance for empirical measures on the real line \cite{Peyre2019}.
Put clearly, we are not concerned with finding the map $T$ but rather applying it and thus obtaining the output $\bm{y}$.
This fact is of notice because this paper is organized such to present two algorithms by design rather than by analysis.

Here we consider a histogram to be a tuple composed of unique values and their respective counts.
(Which essentially is an attribute-value pair representation of a histogram. The attribute is datum that appears at least once, and the value is its corresponding frequency.)
We may use unique values and counts independently.
Both are connected due to the fact that $\unique(\cdot)$ deliberately returns sorted unique values and $\counts(\cdot)$ returns the counts also for sorted unique values.
In addition, observe that a sorted vector may be constructed by concatenating groups of its sorted unique values.
In the next subsection, we further this observation.

\subsection{Group Mapping Law}

The group mapping law \cite{Zhang1992} is an insightful contribution in image processing to provide integer transformation functions for histogram specification.
It does so by minimizing the $\ell^{1}$ norm of the distance between histograms while considering group mappings.
The original setting only considers unsigned integer transformations.
It allows algorithms based on lookup tables to provide solutions with linear time complexity.
However, it strongly requires \textit{a priori} knowledge of the set of values in the input and the output, which is not the case for tabular data.

We take up the work by Zhang \cite{Zhang1992} to extend it to the context of ordered assignment solutions.
This extension of the group mapping law will be useful to minimize the approximation error to the reference by representing the sorted output, $\sort(\bm{y})$, in function of its unique values.

\begin{definition}
    We define the group mapping law matrix $\bm{A}$ as the matrix of size $n \times m$ of the form $\bm{A}=[(\bm{A})_{0 \dots n-1, 0}, (\bm{A})_{0 \dots n-1, 1}, \dots, (\bm{A})_{0 \dots n-1, m-1}]$, where each column vector $(\bm{A})_{0 \dots n-1, j}$ is defined as $(\bm{A})_{\jjj{\omega}, j}=[1, \dots, 1]^{\top}$ and zeros elsewhere, for $j=0 \dots m-1$, where $\bm{\omega}=\cumsum([0, \psi_{0}, \psi_{1}, \dots, \psi_{m-1}]^{\top})$ and $\bm{\psi}=\counts(\bm{x})$.
    Appropriately, the matrix $\bm{A}$ is designed such that $\bm{Ae}:=x_{\bm{\phi}}$, where $\bm{e}=\unique(\bm{x})$ and $\bm{\phi}=\argsort(\bm{x})$.
\end{definition}

Accordingly,
\begin{equation}\label{eq:bmA}
    \bm{A}=
    \begin{blockarray}{ccccccc}
        \begin{block}{[cccccc]c}
            1      &        &  &        &  &        & \mathsmaller{\omega_{0}}   \\
            \vdots &        &  &        &  &        & \mathsmaller{\vdots}       \\
            1      &        &  &        &  &        & \mathsmaller{\omega_{1}-1} \\
                   & 1      &  &        &  &        & \mathsmaller{\omega_{1}}   \\
                   & \vdots &  &        &  &        & \mathsmaller{\vdots}       \\
                   & 1      &  &        &  &        & \mathsmaller{\omega_{2}-1} \\
                   &        &  &        &  &        &                            \\
                   &        &  & \ddots &  &        & \mathsmaller{\vdots}       \\
                   &        &  &        &  &        &                            \\
                   &        &  &        &  & 1      & \mathsmaller{\omega_{m-1}} \\
                   &        &  &        &  & \vdots & \mathsmaller{\vdots}       \\
                   &        &  &        &  & 1      & \mathsmaller{\omega_{m}-1} \\
        \end{block}
        \mathsmaller{0} & \mathsmaller{1} & & \mathsmaller{\cdots} & & \mathsmaller{m-1} & \\
    \end{blockarray}.
\end{equation}
In \eqref{eq:bmA}, labels below mark the columns of $\bm{A}$, and labels to the right mark the rows of $\bm{A}$.
Omitted elements are zeros whereas $[1, \dots, 1]^{\top}$ represents all ones.
Each group of ones, indexed by slices of the form $\jjj{\omega}$, has $\psi_{j}$ elements.
Notice that $\omega_{0}=0$ and that $\omega_{m}-1=n-1$.

\begin{algorithm}[t]
    \DontPrintSemicolon
    \KwData{$\bm{x}$, $\bm{v}$, $p$}
    \KwResult{$\bm{y}$}
    \Begin{
        \tcp*[h]{Ancillary counts and group indices arrays.}\;
        $\bm{\psi} \leftarrow \counts(\bm{x})$\;
        $\bm{\omega} \leftarrow \cumsum([0, \psi_{0}, \psi_{1}, \dots, \psi_{m-1}]^{\top})$\;
        \tcp*[h]{Optimal unique values.}\;
        \For{$j=0$ \KwTo $m-1$}{
            \tcp*[h]{Fréchet $p$-mean \cite{Frechet1948}.}\;
            $u_{j} \leftarrow \argmin_{u_{j}} \norm{u_{j}-v_{\jjj{\omega}}}_{p}$
        }
        \tcp*[h]{Reconstruction by ordered assignment \cite{Karp1975, Werman1986}.}\;
        $\bm{\phi} \leftarrow \argsort(\bm{x})$\;
        $y_{\bm{\phi}} \leftarrow \bm{Au}$\;
        \Return{$\bm{y}$}
    }
    \caption{Histogram Specification by Assignment of Optimal Unique Values\label{algo:hs}}
\end{algorithm}

The group mapping law matrix $\bm{A}$ is a sparse rectangular matrix, with $n$ nonzero values.
For $n=m$, it is the identity matrix.
It is associated with an input $\bm{x}$.
The slice $(\bm{Ae})_{\jjj{\omega}}=[e_{j}, \dots, e_{j}]^{\top}$ corresponds to the group of length $\psi_{j}$ composed of the unique value $e_{j}$ repeatedly.
To provide a clearer illustration of $\bm{\omega}$, consider the following.
The sorted values of $\bm{x}$, $x_{\bm{\phi}}$, can be constructed by repeating $\psi_{j}$ times each unique element $e_{j}$, for $j=0 \dots m-1$, followed by concatenating all $m$ groups.
The indices that mark the start of each group are given by $\bm{\omega}=\cumsum([0, \psi_{0}, \psi_{1}, \dots, \psi_{m-1}]^{\top})$, which we can then use to specify index ranges, as follows.
\begin{equation}
    \bm{Ae}:=[\underbrace{e_{0}, \dots, e_{0}}_{(\bm{Ae})_{\omega_{0} \dots \omega_{1}-1}}, \underbrace{e_{1}, \dots, e_{1}}_{(\bm{Ae})_{\omega_{1} \dots \omega_{2}-1}}, \dots, \underbrace{e_{m-1}, \dots, e_{m-1}}_{(\bm{Ae})_{\omega_{m-1} \dots \omega_{m}-1}}]^{\top}
\end{equation}

\subsection{Ordered Assignment}

The solution to the assignment problem by ordered assignment is as follows \cite{Karp1975, Werman1986, Peyre2019}.
Given an input $\bm{x}\in\mathcal{X}^{n}$ and a reference $\bm{v}\in\mathcal{V}^{n}$, the solution $\bm{y}\in\mathcal{Y}^{n}$ to the assignment problem is $y_{\bm{\phi}}=\sort(\bm{v})$, where $\bm{\phi}=\argsort(\bm{x})$.
(To abridge notation, we assume hereafter that $\bm{v}$ is already sorted and is strictly nondecreasing \cite{Karp1975}, and also that $\mathcal{V}=\mathbb{R}$ and thus $\mathcal{Y}=\mathbb{R}$ as previously noted in \autoref{tab:definitions}.)

However, this solution may not guarantee mapping bijectivity and therefore may yield conflicting assignments.
Ordered assignment solutions are bijective if and only if $\counts(\bm{x})=[1, \dots, 1]^{\top}$.
That is, if and only if $\bm{x}$ is an array of all different values.
The pitfalls of conflicting assignments can manifest as shearing in image processing, or inconsistent transformation in tabular data (equal, independent observations being mapped to different values).
In literature, to apply the ordered assignment solution for exact histogram specification, Coltuc \textit{et al.} \cite{Coltuc2006} uses local information to obtain a strict ordering of values.
It should be clear that this procedure cannot apply to data that has no local structure.

To provide a conflict-free assignment, we must map each group of unique values individually.
Thus, we reconstruct the output $\bm{y}$ by ordered assignment of the sorted array of the groups of unique values $\bm{Au}$.
In other words, we have $y_{\bm{\phi}}=\bm{Au}$, where $\bm{\phi}=\argsort(\bm{x})$.

\subsection{Optimal Unique Values}

Nevertheless, it remains that we find the optimal unique values.
To that end, we propose obtaining the least $p$-norm approximation of the output unique values w.r.t. the reference.
We use the proposed extension of the group mapping law, the matrix $\bm{A}$, to formulate the minimization of the $\ell^{p}$ norm of a linear function of the output unique values, as follows.
\begin{equation}\label{eq:argminbmu}
    \argmin_{\bm{u}} \norm{\bm{Au}-\bm{v}}_{p},
\end{equation}
where $\bm{u}$ is the array of unique values of the output, $\bm{v}$ is the reference, and $\bm{A}$ is the group mapping law matrix of $\bm{x}$.

For instance, in \eqref{eq:argminbmu}, for $p=2$, we have the least-square solution $\bm{u}=\bm{A}^{+} \bm{v}$, where $\bm{A}^{+}$ is the Moore–Penrose inverse of $\bm{A}$.
Each row vector $(\bm{A}^{+})_{j, 0 \dots n-1}$ of $\bm{A}^{+}$ is of the form $(\bm{A}^{+})_{j, \jjj{\omega}}=[1/{\psi_{j}}, \dots, 1/{\psi_{j}}]$ and zeros elsewhere, for $j=0 \dots m-1$.
$\bm{A}^{+}$ is as sparse as $\bm{A}$.
This closed-form solution is of notice because we can gather that each group has solutions that are local to their group.
In this case, the solution is that each group of unique values is the mean of the same corresponding slices of the reference.
Further, for $p=1$, this problem is known to be an instance of linear programming (LP) \cite{Kantorovich2006}.
However, neither the Moore-Penrose inverse nor the LP solutions are suitable for large-scale problems.
In the next section, we detail on how to solve $\eqref{eq:argminbmu}$ taking into consideration the structure of the matrix $\bm{A}$.

\section{Algorithms}

We can now present the two algorithms.
The first, \autoref{algo:hs}, presents a fast algorithm for group histogram specification.
For $p\in\set{1, 2, \infty}$, it has closed-form solutions.
The second, \autoref{algo:fqt}, presents a vectorized algorithm for the quantile transformation problem.
They both work similarly.
Firstly, we obtain the counts array $\bm{\psi}$ and calculate the ancillary group indices array $\bm{\omega}$.
Secondly, we obtain the optimal unique values $\bm{u}$.
Lastly, we reconstruct the output $\bm{y}$ from the optimal unique values by ordered assignment.
In the following subsections, we detail the obtainment of the optimal unique values for each case.

\subsection{Fast Barycenters on the Real Line (\autoref{algo:hs})}

\begin{algorithm}[t]
    \DontPrintSemicolon
    \KwData{$\bm{x}$, $\alpha \leftarrow 0$, $\beta \leftarrow 0$}
    \KwResult{$\bm{y}$}
    \Begin{
        $\bm{\psi} \leftarrow \counts(\bm{x})$\;
        $\bm{\omega} \leftarrow \cumsum([0, \psi_{0}, \psi_{1}, \dots, \psi_{m-1}]^{\top})$\;
        $\gamma \leftarrow 1/{(n+1-\alpha-\beta)}$\;
        \tcp*[h]{Chebyshev approximation \cite{Iske2018}.}\;
        $\bm{u} \leftarrow \gamma(\omega_{0 \dots m-1}+\omega_{1 \dots m}+1-2\alpha)/{2}$\;
        $\bm{\phi} \leftarrow \argsort(\bm{x})$\;
        $y_{\bm{\phi}} \leftarrow \bm{Au}$\;
        \Return{$\bm{y}$}
    }
    \caption{Fast Quantile Transformer (Vectorized)\label{algo:fqt}}
\end{algorithm}

We take up the problem posed in \eqref{eq:argminbmu}.
The problem consists in finding, given reference $\bm{v}$, the optimal $\bm{u}$ in
\begin{equation}
    \argmin_{\bm{u}} \bigl(\sum_{i=0}^{n-1}\abs{(\bm{Au})_{i}-v_{i}}^{p}\bigr)^{1/{p}}.
\end{equation}
Recall that, by definition, $(\bm{Au})_{\jjj{\omega}}=[u_{j}, \dots, u_{j}]^{\top}$.
Then, the minimization functional can be rewritten as
\begin{equation}
    \argmin_{\bm{u}} \bigl(\sum_{j=0}^{m-1}\sum_{k=0}^{\psi_{j}-1}\abs{u_{j}-(v_{\jjj{\omega}})_{k}}^{p}\bigr)^{1/{p}},
\end{equation}
which is equivalent to $m$ scalar optimization problems.
We shall analyze the $j$th term without loss of generality.
\begin{equation}\label{eq:argminuj}
    \argmin_{u_{j}} \norm{u_{j}-v_{\jjj{\omega}}}_{p}
\end{equation}
In \eqref{eq:argminuj}, we have a $p$-barycenter on the real line, or more properly, a Fréchet $p$-mean \cite{Frechet1948}.
Its existence is well-defined for the values of $p$ considered, $p \ge 1$.
Also, it is unique on the real line \cite[Section 3.2]{Iske2018}.
There are three notable results associated with the Fréchet $p$-mean on the real line.
For $p=1$, it is the median, and for $p=2$, it is the arithmetic mean \cite{Frechet1948}.
The last notable result is for $p=\infty$, where we have the best approximation with respect to the Chebyshev norm, that is, the point the furthest inside the feasible set \cite[Corollary 5.2]{Iske2018}.
In this case, the set is the one defined by the convex hull of $\mathcal{U}_{j}=\set{v_{\omega_{j}}, \dots, v_{\omega_\jj}}$.
On the real line, it is the midpoint of such interval, $u_{j}=(\min\mathcal{U}_{j}+\max\mathcal{U}_{j})/{2}$.
Since $\bm{v}$ is sorted, it is even simpler: $u_{j}=(v_{\omega_{j}}+v_{\omega_\jj})/{2}$.

The three closed-form solutions to \eqref{eq:argminuj} reveal that \autoref{algo:hs} is fast for $p\in\set{1, 2, \infty}$.
The minimization problem in \eqref{eq:argminuj} can be easily found by scalar optimization otherwise.
The complete method appears in \autoref{algo:hs}.

\subsection{Quantile Transformation (\autoref{algo:fqt})}

We can further use the histogram specification algorithm presented in \autoref{algo:hs} to transform data into quantiles by specifying the sample CDF to approximate the CDF of a uniform distribution.
For this purpose, and to align with computational statistics literature \cite[Table 3]{Hyndman1996}, in \autoref{algo:fqt}, we accommodate the $\alpha$ and $\beta$ interpolation parameters defined by Hyndman and Fan \cite{Hyndman1996} in the definition of our reference $\bm{v}$, as follows.
\begin{equation}
    v_{i}=\gamma(i+1-\alpha), \text{ where } \gamma=1/{(n+1-\alpha-\beta)}.
\end{equation}
In \autoref{algo:fqt}, the default values included for $\alpha$ and $\beta$ refer to Type 6 interpolation parameters \cite{Hyndman1996}.

In addition, it is easy to verify that any sorted array that is symmetric along its midpoint has the same barycenter for all $p \ge 1$.
For instance, the median, mean, and the midpoint all have the same value.
This is the case in uniform distributions.
For a uniform reference, any slice $v_{\jjj{\omega}}$ finds the same minimizer in \eqref{eq:argminuj} for all $p \ge 1$.
Therefore, in \autoref{algo:fqt}, we adopt the midpoint closed-form solution to \eqref{algo:fqt}, as it is vectorizable with few operations.

\begin{table*}
    \centering
    \caption{Approximation Error for Column-Wise Histogram Specification of Tabular Data Sets}
    \label{tab:results}
    \begin{tabular}{@{}ccccccc@{}}
        \toprule
        \multirow{2}{*}{Data}          & \multirow{2}{*}{Reference} & \multirow{2}{*}{$p$} & \multicolumn{4}{c}{Method}                                                                                                                                                                \\\cmidrule(lr){4-7}
                                       &                            &                      & \multicolumn{1}{c}{Estimation \cite{Hyndman1996}} & \multicolumn{1}{c}{Approximation \cite{Dunning2019}} & \multicolumn{1}{c}{\autoref{algo:fqt}} & \multicolumn{1}{c}{\autoref{algo:hs}} \\\midrule
        \multirow{6}{*}{Breast Cancer} & \multirow{3}{*}{Uniform}   & 1                    & 17.747                                            & 14.473                                               & 3.396                                  & 3.396                                 \\
                                       &                            & 2                    & 0.197                                             & 0.146                                                & 0.082                                  & 0.082                                 \\
                                       &                            & $\infty$             & 0.023                                             & 0.011                                                & 0.011                                  & 0.011                                 \\
                                       & \multirow{3}{*}{Normal}    & 1                    & 476.517                                           & 105.829                                              & 26.891                                 & 26.891                                \\
                                       &                            & 2                    & 30.706                                            & 3.076                                                & 2.439                                  & 2.363                                 \\
                                       &                            & $\infty$             & 3.200                                             & 0.643                                                & 0.671                                  & 0.460                                 \\
        \multirow{6}{*}{Diabetes}      & \multirow{3}{*}{Uniform}   & 1                    & 148.137                                           & 89.303                                               & 88.711                                 & 88.711                                \\
                                       &                            & 2                    & 6.256                                             & 3.314                                                & 3.314                                  & 3.314                                 \\
                                       &                            & $\infty$             & 0.530                                             & 0.265                                                & 0.264                                  & 0.264                                 \\
                                       & \multirow{3}{*}{Normal}    & 1                    & 2237.136                                          & 339.258                                              & 329.773                                & 329.773                               \\
                                       &                            & 2                    & 95.986                                            & 13.569                                               & 13.543                                 & 13.295                                \\
                                       &                            & $\infty$             & 5.276                                             & 2.214                                                & 2.216                                  & 1.458                                 \\
        \multirow{6}{*}{Iris}          & \multirow{3}{*}{Uniform}   & 1                    & 9.978                                             & 8.808                                                & 8.662                                  & 8.662                                 \\
                                       &                            & 2                    & 0.578                                             & 0.525                                                & 0.523                                  & 0.523                                 \\
                                       &                            & $\infty$             & 0.106                                             & 0.095                                                & 0.093                                  & 0.093                                 \\
                                       & \multirow{3}{*}{Normal}    & 1                    & 80.099                                            & 37.102                                               & 34.334                                 & 34.334                                \\
                                       &                            & 2                    & 11.169                                            & 2.339                                                & 2.244                                  & 2.226                                 \\
                                       &                            & $\infty$             & 3.362                                             & 0.627                                                & 0.639                                  & 0.499                                 \\
        \multirow{6}{*}{Wine}          & \multirow{3}{*}{Uniform}   & 1                    & 13.131                                            & 10.507                                               & 8.994                                  & 8.994                                 \\
                                       &                            & 2                    & 0.381                                             & 0.329                                                & 0.319                                  & 0.319                                 \\
                                       &                            & $\infty$             & 0.041                                             & 0.040                                                & 0.039                                  & 0.039                                 \\
                                       & \multirow{3}{*}{Normal}    & 1                    & 134.022                                           & 51.738                                               & 33.782                                 & 33.782                                \\
                                       &                            & 2                    & 13.811                                            & 1.908                                                & 1.252                                  & 1.250                                 \\
                                       &                            & $\infty$             & 2.662                                             & 0.232                                                & 0.221                                  & 0.186                                 \\
        \bottomrule
    \end{tabular}
\end{table*}

\section{Results}

In order to validate the proposed algorithms, we include two sets of results from numerical experiments that demonstrate certain properties of the proposed algorithms.
The first set of results regards the task of the histogram specification of tabular data sets.
There, we verify how our method fares when compared to traditional sample quantile estimation and state-of-the-art approximate quantile computation methods with respect to least $p$-norm histogram specification.
The second set of results regards the task of exact histogram specification of images.
We show that, while local exact histogram specification methods provide a better approximation of the histogram of an output with respect to a reference, these methods generate artifacts that ours is designed not to.
Specifically, our method provides the least $p$-norm solution while preserving the mapping bijectivity.

\subsection{Histogram Specification of Tabular Data Sets}

In this first set of results, we evaluated the $\ell^{p}$ norm of the approximation error for the task of column-wise histogram specification of tabular data sets with respect to uniform and normal references.
We considered the popular Breast Cancer, Fisher's Iris, and Wine tabular data sets, all available on the UCI Machine Learning Repository \cite{Dua2019}.
The likewise popular Diabetes data set is by Efron \textit{et al.} \cite{Efron2004}.
The first baseline refers to traditional sample quantile estimation by the method of Hyndman and Fan \cite{Hyndman1996}.
We estimate $n$ sample quantiles, then, we transform the input data by evaluating the interpolated CDF estimate given by the quantiles.
The second baseline refers to state-of-the-art approximate quantile computation by the method of Dunning and Ertl \cite{Dunning2019}.
We batch update a $t$-digest data structure with the input data, then, we transform the input data by evaluating the approximate CDF given by the $t$-digest.
To specify references other than uniform, we evaluate the uniformly transformed data---interpreted as quantiles---using the inverse CDF of the reference (see, e.g., the inversion method \cite[Theorem 2.1]{Devroye1986}).
We must use the inversion method to specify a normal reference for the estimation and approximation baselines, and for \autoref{algo:fqt}, as these only output quantiles.
In \autoref{tab:results}, we present these results.

The proposed method of histogram specification by assignment of optimal unique values (\autoref{algo:hs}) presented the least error in all configurations.
This particular algorithm directly optimizes the approximation error, thus, that is what we expected.
The proposed fast quantile transformer (\autoref{algo:fqt}) also presented supporting results.
For a uniform reference, \autoref{algo:fqt} presents the same approximation error as \autoref{algo:hs}.
That is due to the previously discussed result that, for uniform references, the best $p$-norm approximation has the same value regardless of $p$.
For a normal reference, \autoref{algo:fqt} presents the same approximation error as \autoref{algo:hs} only for $p=1$.
That is because the inversion method only minimizes the $\ell^{1}$ norm.
For the Breast Cancer and Iris data sets, the approximate quantile computation baseline presents an approximation error lower than that of \autoref{algo:fqt}.
But for the Diabetes and Wine data sets, it presents higher error.
We deem this behavior due to the nature of approximate computation.
As the objective is to provide accurate quantile computation with bounds relative to a particular measure, it may unintentionally provide better results in certain configurations.
Moreover, the time complexity of the estimation baseline is $O(n \log n+m^{2})$ (i.e., quadratic in number of unique values) and the approximation baseline is $O(n \log^{2} n)$, whereas \autoref{algo:hs} is $O(n \log n)$ for $p\in\set{1, 2, \infty}$ and \autoref{algo:fqt} is always $O(n \log n)$.

\begin{figure*}
    \centering
    \includegraphics[scale=1]{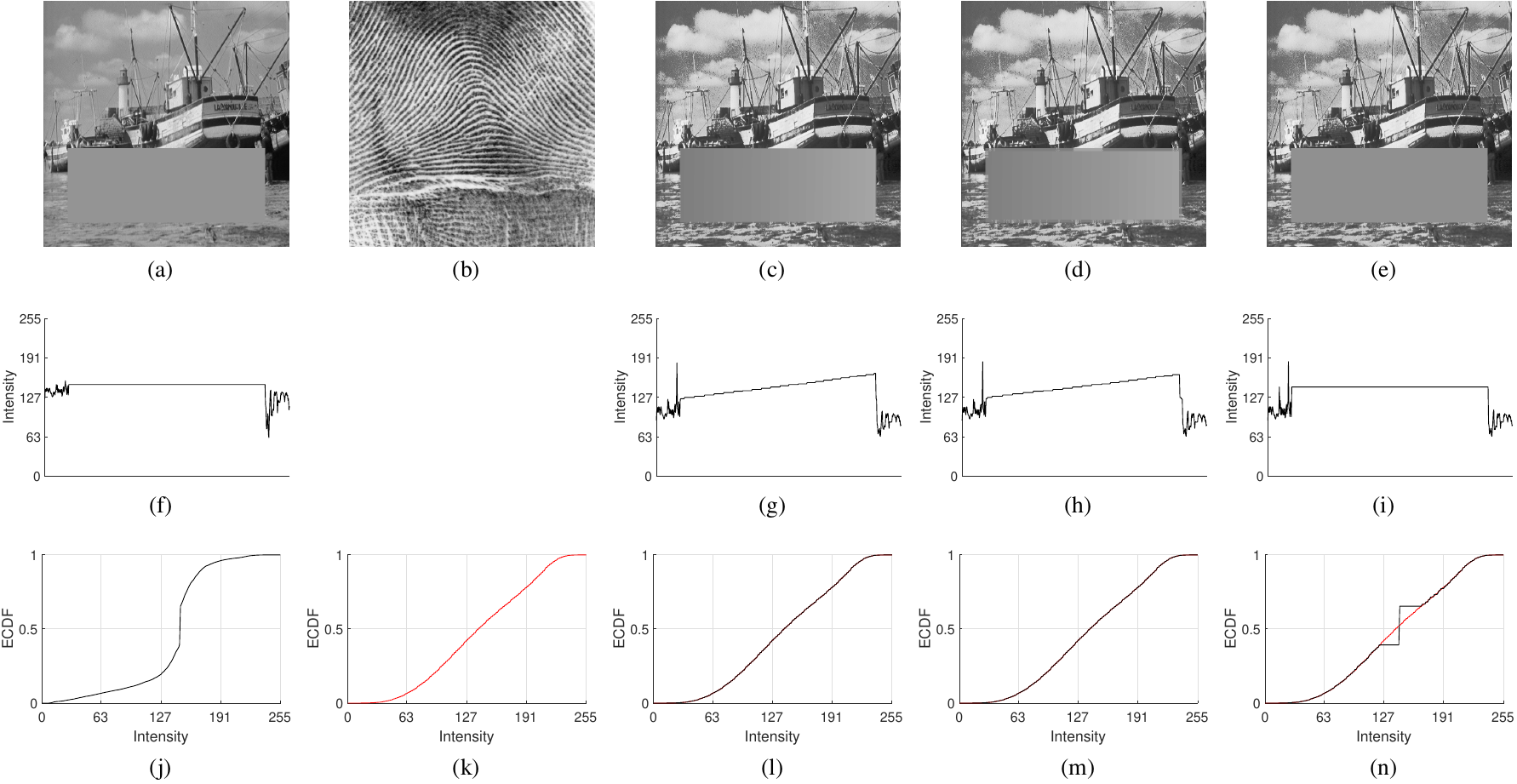}
    \caption{An exact histogram specification example.
        First, we inscribed a constant-valued rectangle to simulate, e.g., a mixed content raster graphic.
        Then, we performed the exact histogram specification of input image (a), using reference image (b), obtaining output images (c)--(e).
        Next, we plotted scan lines (f)--(i) along the columns of the middle of the inscribed rectangle in input (a) and output images (c)--(e), respectively, to evidence the noticeable gradient artifact in (c), (d), and lack thereof in (a), (e).
        After that, we plot empirical CDFs (ECDFs) in (j)--(n) corresponding to the empirical distribution of intensity values in images (a)--(e).
        In (j), the jump at intensity value 148 refers to the large constant-valued area of the inscribed rectangle.
        In (l), (m), we show the ECDF of local exact histogram specification methods.
        Although local exact histogram specification methods faithfully transport the ECDF of the reference, it introduces undesirable artifacts.
        In (n), the jump at intensity value 144 also refers to the inscribed rectangle.
        This alludes to the fact that the constant-valued rectangle remains of constant value, i.e. that its origin intensity of 148 was mapped to an output intensity of 144 such that it best approximates the reference ECDF (in red).
        (a) Input image (boat).
        (b) Reference image (fingerprint).
        (c)--(e) Output images.
        (c) Coltuc \textit{et al.} \cite{Coltuc2006}.
        (d) Nikolova and Steidl \cite{Nikolova2014}.
        (e) Ours (\autoref{algo:hs}).
        (f)--(i) Scan lines along the columns of the middle of the inscribed constant-valued rectangle in corresponding images (a), (c)--(e).
        (j)--(n) ECDF of the intensity values of corresponding images (a)--(e)%
    }
    \label{fig:1}
\end{figure*}

\subsubsection{Limitations}

An important limitation is that, while approximate quantile computation algorithms generally support streaming data use cases \cite{Chen2020}, the algorithms proposed in this paper do not.

\subsection{Exact Histogram Specification}

In this second set of results, we demonstrate the effectiveness of the proposed method for the task of exact histogram specification.

In \cite{Coltuc2006}, Coltuc \textit{et al.} acknowledge that their approach of strict ordering by leveraging local information is principled for natural images.
In \cite{Nikolova2014}, Nikolova and Steidl acknowledge that using local information may generate artifacts.
They propose simple heuristics such as masking large flat areas in an input image before histogram specification.

However, the assumption that all images are natural or that masks are available does not hold in content-agnostic systems such as electronic visual displays.

Further, suppose we are to preprocess medical imaging data to enhance visualization for an electronic visual display.
We must be careful not to preprocess data in a way that contextually local information in carefully reconstructed data is altered.
The same applies in physical sciences, in general.
Two pixels, neighboring or not, that have the same value refer to two independent physical measurements that yielded the same value.
Processing them---even if for the purpose of image enhancement---in a way that maps them to different values may scientifically invalidate their meaning.

In Fig. \ref{fig:1}, we exhibit a critical artifact of local exact histogram specification.
To reproduce the artifact, we inscribe a flat (constant-valued) rectangle in a natural grayscale image.
This represents imagery common to any configuration wherein natural images are interwoven with computer graphics-generated images, such as synthetic vision systems; broadcast and multimedia video feeds; and graphical elements in user interfaces.

We compare our method to Coltuc \textit{et al.} \cite{Coltuc2006}, a seminal work on the subject of exact histogram specification, and to Nikolova and Steidl \cite{Nikolova2014}, a state-of-the-art method for the same.
We use two standard testing images, specifically, the boat and fingerprint images.
We apply the method of Coltuc \textit{et al.} \cite{Coltuc2006} and Nikolova and Steidl \cite{Nikolova2014} using the boat image for the input and the fingerprint image for the reference.
To apply \autoref{algo:hs}, for each image we concatenate all columns to obtain a single column vector containing all pixels, and additionally we set $p=1$.
After processing, we reshape the output column vector back to its image form.
The results are included in Fig. \ref{fig:1} (c)--(e).

While exact histogram specification does, in fact, produce total transformation of an input image such that its histogram is specified by a reference histogram, it generates gradient artifacts due to failure of obtaining a strict ordering of pixels and applying the ordered assignment solution with stable sorting \cite{Knuth1998}.
Should unstable sorting be employed, we would see noise instead of a gradient.
The fact, however, is that an entire constant-valued region was mapped to a great number of different values.

As observable in Fig. \ref{fig:1} (n), our method minimizes the distance between the output histogram (in black) and the reference histogram (in red) while preserving bijectivity.
Note that the discontinuity seen in Fig. \ref{fig:1} (j), (n) at intensities 148 and 144, respectively, is due to the contribution of intensity values of the inscribed rectangle to the empirical distribution.
Our approach correctly maps same-intensity values group-wise such that the output histogram best approximates the reference histogram in $p$-norm.
Refer additionally to \autoref{fig:1} for an extensive description of each subfigure.

In future work, it may also be possible to propose heuristics to separate computer-generated regions from natural image regions and propose a solution combining our method to state-of-the-art local exact histogram specification.

\section{Conclusion}

The main goal of this work was to provide applied data scientists with a novel histogram specification method that is both computationally practical and numerically sound.
In short, we extended the group mapping law, included it in a convex formulation to optimize for the best unique values, and proposed reconstructing the output by assignment of optimal unique values.
Then, we proposed a general algorithm for any reference and a fast algorithm for uniform references.
The former is also fast for $p\in\set{1, 2, \infty}$.
The first algorithm is useful for transforming the histogram of the input into any given reference, while the second is useful for data normalization---a common practice in data preprocessing.
The decrease in complexity provided now allows tractability for the exact column-wise transformation of large-scale tabular data.
The results demonstrated desirable qualities of the proposed method, such as the ability to perform least $p$-norm histogram specification of tabular data, and the ability to perform exact histogram specification without compromising the bijectivity of the intensity transformation.
Future works can consider other distance measures between sample CDF spaces, and re-cast this problem in the computational optimal transport framework (e.g., regularized transport solutions can be useful for outlier smoothing).

\bibliographystyle{IEEEtran}
\bibliography{ms}

\end{document}